\documentstyle[prl,aps,twocolumn]{revtex}
\begin{document}
\twocolumn[\hsize\textwidth\columnwidth\hsize\csname@twocolumnfalse%
\endcsname
\title{Sigma models as perturbed conformal field theories}
\author{Paul Fendley}
\address{Physics Department, University of Virginia,
Charlottesville VA 22901}
\date{October 4, 1999}
\maketitle
\begin{abstract}

We show that two-dimensional sigma models are related to certain
perturbed conformal field theories. When the fields in the sigma model
take values in a space $G/H$ for a group $G$ and a maximal subgroup
$H$, we argue that the corresponding conformal field theory is the
$k\to\infty$ limit of the coset model $(G/H)_k$, and the perturbation
is related to the currents of $G$.  Non-perturbative
instanton contributions to the sigma model free energy are
perturbative when $k$ is finite.  We use this mapping
to find the free energy for the ``$O(n)$'' ($=O(n)/O(n$--$1)$)
sigma model at non-zero temperature.  It also results in a new
approach to the $CP^{n}$ model.

\end{abstract}
\pacs{PACS numbers: ???}  ]  

Sigma models are used frequently in particle physics and condensed-matter
physics to describe Goldstone excitations and their interactions.
When a field transforming under some symmetry group $G$ has an
expectation value invariant under some subgroup $H$, the low-energy
modes of the field take values in the manifold $G/H$. The $G/H$ sigma
model is the field theory
describing these low-energy modes.  Even
in two dimensions, where quantum effects restore the original symmetry
group $G$ and the low-energy excitations are massive, sigma models are
very useful. Two-dimensional sigma models have been the subject of a
huge amount of study because they can be interesting toy models for
gauge theories, because they often arise in experimentally-realizable
condensed-matter systems, because this is the highest dimension in
which they are naively renormalizable, and because of the powerful
theoretical methods applicable.

One of the great breakthroughs in two-dimensional field
theory was the realization that many known models (and even more
previously-unknown theories) could be written as perturbed conformal
field theories \cite{Zam}. One starts with the model
at its critical point, which is described by a conformal field
theory. In many conformal field theories, all the relevant operators
are known. One can thus define a massive field theory by adding some
relevant operator to the action.  This defines the theory to all
orders in perturbation theory, even if the action of the conformal
field theory is not known. Not any such model is integrable (most are
not) but if it is, one can apply a variety of techniques to
find for example the exact $S$ matrix and the free
energy. Literally dozens of infinite hierarchies of models have been
solved over the last decade, the most famous single model being the
Ising field theory at $T=T_c$ in a magnetic field \cite{Zam}.

Sigma models have stood somewhat apart from this line of
development. Some exact $S$ matrices (for example, in the ``$O(n)$''
model \cite{ZamZam} and the principal chiral model \cite{Wieg}) have
been known for quite some time.  The energy in a magnetic field at
zero temperature in these models can be computed \cite{Weiner,Hasen},
but further progress has been slow. Usually one computes
finite-temperature properties in an integrable model using the
thermodynamic Bethe ansatz, but often it is not known how to
categorize the solutions of the Bethe ansatz equations, a necessary
step for the thermodynamics. There are a few cases where the
computation is possible: the ``$O(3)$'' model and its sausage
deformation \cite{Tsvel,FZ}, the ``$O(4)$'' model (equivalent to the $SU(2)$
principal chiral model) \cite{Fateev} and the supersymmetric $CP^n$
models \cite{FI}. All these models have an intriguing similarity: they
can all be expressed as limits of certain perturbed conformal field
theories \cite{Poly,FZ,Fateev,FI}. The purpose of this paper is to
explain the general principle behind these results, and to extend it
further.

We will make a general conjecture that these $G/H$ sigma models are
equivalent to the $k\to\infty$ limit of a particular perturbation of
the coset conformal field theory $(G/H)_k$. The utility of this result
is threefold. First of all, it uncovers a nice general structure of
$G/H$ sigma models. Moreover, it makes it possible to use the powerful
methods of perturbed rational conformal field theory on sigma models.
Finally, a great deal is known about integrable perturbed conformal
field theory (much more than in sigma models), and therefore these
results can be applied to sigma models. For example, technical
complications had prevented the computation of the free energy of the
``$O(n)$'' sigma models directly. We will show how this
construction enables this computation for any $n$.

One of the interesting consequences of this reformulation is that
non-perturbative instanton contributions to the free energy are
perturbative when $k$ is finite \cite{FZ}. For example, in the appropriate
perturbation of the coset models $(O(3)/O(2))_k$, there is a
contribution to the free energy at order $k$ times any integer. When
one takes $k\to\infty$ to obtain the ``$O(3)$'' sigma model, this
contribution turns into the instanton contribution, which is not
polynomial in the perturbing parameter. This remains true
even when a theta term is
present. For example, this yields the result that the $SU(N)/SO(N)$
sigma models are integrable when $\theta=\pi$, and they flow to the
$SU(N)_1$ conformal field theory \cite{theta}.

We study symmetric spaces $G/H$, where $G$ and $H$ are Lie groups,
and $H$ is a maximal subgroup of $G$. 
The $G/H$ sigma model has action
\begin{equation}
S=\int d^2z\ g_{ij}(X)\partial_\mu X^i(z,\overline{z}) \partial^\mu
X^j(z,\overline{z}),
\label{sig}
\end{equation}
where $z$ and $\bar z$ are coordinates for two-dimensional spacetime,
and $X^i$ and $g_{ij}(X)$ are the coordinates and metric for the manifold
$G/H$. Symmetric spaces have non-vanishing curvature, so (\ref{sig}) defines
an asymptotically-free massive field theory.  When $G={H} \times H$
and $H$ is a simple Lie group diagonally embedded in $G$, the
resulting sigma model is called the principal chiral model.  Another
example is the ``$O(n)$'' model, where $G=O(n)$ and $H=O(n-1)$. This
space is an $n-1$ dimensional sphere: $O(n)$ is the
rotational symmetry of the sphere, while $O(n-1)$ is the subgroup
leaving a given point fixed.

A coset conformal field theory utilizes the affine Kac-Moody algebra $G_k$
defined by the operator product
\begin{equation}
J^A(z) J^B(w) = {k\over (z-w)^2} + {f^{ABC} J^C(w)\over z-w}+\dots, $$
\label{OPE}
\end{equation}
where the $f^{ABC}$ are the structure constants of the ordinary Lie
algebra for $G$ and $k$ is called the level; $k$ is a positive integer
for a compact Lie group. A conformal field theory with current algebra
$G_k$ is called a Wess-Zumino-Witten model, and is equivalent to the
principal chiral model for $G$ plus an extra piece called the
Wess-Zumino term \cite{WZW}. The central charge (coefficient of the
conformal anomaly) of the $G_k$ WZW model is $k
\,\hbox{dim}G/(k+\tilde h)$, where $f^{ACD}f^{BCD} = \tilde h
\delta_{AB}/2$. For $G=SU(n)$, $\tilde h=n$, while for $G=SO(n)$,
$\tilde h=n$--$2$ (for $n\ge 4)$. The primary fields of the WZW model
have scaling dimensions $x_j=2C_j/(k+\tilde h)$, where $C_j$ is the
quadratic Casimir defined by $T^A T^A= C_j {\cal I}$, with the $T^A$
the generators of the Lie algebra of $G$ in the $j$th representation
and ${\cal I}$ the identity matrix. All the other scaling fields arise
from the operator product of the $J^A(z)$ with the primary fields; it
follows from (\ref{OPE}) that $J$ has dimension $1$ and all fields
have dimensions $x_j$ plus an integer.

Given a subgroup $H$ of $G$, a $(G/H)_k$ coset conformal field theory
is defined from the generators of $G_k$ not in the subalgebra $H_l$
($l/k$ is the index of the embedding of $H$ into $G$) \cite{GKO}.  The
central charge of this new conformal field theory is $c_G -
c_H$. The energy-momentum tensor obeys the orthogonal
decomposition $T_G=T_H + T_{G/H}$, so a field $\phi_G$ (some
representation of $G_k$) decomposes into representations $\phi_H^a$ of
$H_l$ as
\begin{equation}
\phi_G = \oplus_a\  \phi^a_{G/H} \otimes \phi^a_H.
\label{decomp}
\end{equation}
The coefficients $\phi^a_{G/H}$ are the fields
of the coset model $(G/H)_k$. A consequence of $G/H$ being
a symmetric space is that the generators of $G$ not in $H$ 
form a real irreducible
representation of $H$ \cite{ZJ}. Thus when the
currents $J^A(z)$ are decomposed into representations of $H_l$, the
resulting fields in the coset model form a real irreducible
representation of $H_l$, which we denote as ${\cal J}^{a}$, for $a =
1\dots,(\hbox{dim}\,G- \hbox{dim}\,H)$.

Obviously, the $G/H$ sigma model cannot be equivalent to a coset
theory $(G/H)_k$, because the latter is massless while the former is
not. A massive field theory is defined by perturbing
$(G/H)_k$ by a relevant operator. We can now state our conjecture
precisely.
\bigskip

\noindent {\it Conjecture} \ \ The sigma model for the symmetric space
$G/H$ is equivalent to the $k\to\infty$ limit of the $(G/H)_k$
coset conformal field theory perturbed by the operator 
\begin{equation}
{\cal O}_{\sigma} \equiv \sum_{a =1}^{\hbox{dim }G-
\hbox{dim }H}
{\cal J}^{a}(z) \overline{\cal J}^{a}(\overline z).
\label{pert}
\end{equation}
\noindent
Because the ${\cal J}^a$ form a real irreducible representation of
$H_l$, their dimension is independent of $a$.
\medskip

This perturbed coset has the general properties
of a sigma model.  In the ultraviolet limit, the perturbation of
$(G/H)_k$ goes away, and its central charge when $k\to\infty$ is dim$G
-$dim$H$. In the ultraviolet limit of the sigma model, asymptotic
freedom means that the manifold $G/H$ becomes flat (e.g.\ in the
$O(n)/O(n$--$1)$ model, the radius of the sphere goes to
infinity). The action (\ref{sig}) reduces to dim$G-$dim$H$ free
bosons, which also have the central charge dim$G-$dim$H$.  Moreover,
when $J^A$ is decomposed into representations of $H_l$, the resulting
field $\phi_H^a$ has dimension going to zero as $k\to\infty$. Thus
the field ${\cal J}^a$ has dimension $1$ in this limit, so the
perturbation ${\cal O}_\sigma$ is of dimension $2$ and so is naively
marginal. It is not exactly marginal -- this is the phenomenon of
dimensional transmutation common to sigma
models.

For principal chiral models, the conjecture is already known to be
true \cite{ABL}, and is reminiscent of an earlier description in terms
of an infinite number of fermion flavors \cite{Poly}.  Since dim$G\,
-$ dim$H=$ dim $H$ here, the perturbation ${\cal O}_\sigma$ of the
coset $H_k \times H_k/H_{2k}$ is in the adjoint of $H_{2k}$. The usual
coset notation for such an operator is $(1,1;\hbox{adjoint})$. This
means that the corresponding $\phi_G$ is a descendant of the identity
primary field in the $H_k$ conformal field theories (i.e.\ $J^A$
operating on the vacuum), and the $\phi_H$ in its decomposition are in
the adjoint of $H_{2k}$.  Such an operator is often called the
``thermal'' operator (because when $k=1$ and $H=SU(2)$, ${\cal
O}_\sigma$ is the thermal operator in the Ising model).
The particles
in the perturbed coset models are kinks 
whose exact $S$ matrices were
conjectured in \cite{ABL,dVF,Gepner}. 
For finite $k$, the kinks form
representations of the quantum-group algebra $U_q(H)$ with
$q=-\exp(i\pi/(k+h))$.  
As $k\to\infty$, $q\to -1$ and
the quantum-group algebra reverts to the ordinary Lie algebra of $H$.
For example, for $SU(4)$, this means that particles are in the $4$,
the $6$ and the $\overline{4}$ representations, giving 14 particles
all together.  Once an ``intertwiner'' is used to change basis, the
$S$ matrices in the $k\to\infty$ limit are those conjectured for the
$H\times H/H$ sigma models in \cite{Wieg}. The exact free energy for
the coset models was found in \cite{Hollo}, using results of
\cite{BR,KNS}.

In the ``$O(3)$'' sigma model, the fields take values on the sphere,
which is the symmetric space $O(3)/O(2)\approx SU(2)/U(1)$. The
curvature (or equivalently, the radius) of the sphere determines the
mass scale of the model. In this case, the conjecture above was put
forth in \cite{FZ}. There it was phrased as taking the $k\to\infty$
limit of the $Z_k$ parafermion theories perturbed by the operator
$\psi_1\overline\psi_1 + h.c.$. Parafermions are a generalization of
fermions which instead pick up $Z_k$ phases when taken around one
another; $\psi_i(z)$ and $\overline\psi_i(\bar z)$ are the
parafermions, where $i$ runs from $1$ to $k-1$. The $Z_k$ parafermion
models can be described by the coset $SU(2)_k/U(1)$, and the operator
${\cal J}^1$ here is indeed the parafermion $\psi_1$, while ${\cal
J}^2=\psi_1^\dagger$ \cite{ZF}.  As opposed to the principal chiral
models, the particles here are in the vector representation of
$O(3)$. Thus our result provides a natural explanation and
generalization of the conjecture of \cite{FZ}.  One interesting thing
about this model is that a topological theta term can be added to the
action. Putting $\theta=\pi$ corresponds to adding the operator
$i{\cal O}_\sigma$ to the action of the coset model \cite{FZ}. The
partition function is still real, because only even powers of the
perturbation appear in the expansion.

We now turn to the general ``$O(n)$'' model, which our conjecture says
should be described by a perturbation of the $O(n)_k/O(n$--$1)_k$
sigma model. The $n$--$1$ fields ${\cal J}^a$ are in the vector
representation of $O(n$--$1)$, which has quadratic Casimir
$(n$--$2)/2$ for $n>4$. Thefore the operator ${\cal O}_\sigma$ is of
dimension $2-(n$--$2)/(k+n-3))$.  To proceed further, we make use of a
level-rank duality which shows that $O(n)_k/O(n$--$1)_k$ coset model
is equivalent to $O(k)_{n-1} \times O(k)_1/ O(k)_n$ \cite{Alt}. In
this dual coset model, the operator ${\cal O}_\sigma$ is
(vector,vector; $1$).  This is precisely the ``electric''-type
perturbation discussed in \cite{Vays}. There it is shown that this
model remains integrable under this perturbation, and that it follows
from the non-local symmetries that the kinks are in the vector
representation of $U_q(O(n))$. (In the thermal perturbation of these
coset models, the particles are in the spinor representations
\cite{Gepner}.)  If we take $k\to\infty$ to reach the sigma model, the
quantum group turns into the ordinary algebra $O(n)$ with the
particles in the vector representation. This agrees with the classic
sigma-model result of \cite{ZamZam}. The $S$ matrices must of course
be the same.

Even though the exact $S$ matrix for these integrable sigma models has
been known for 20 years, the free energy at non-zero temperature has
not been computed. One usually can calculate the free energy of an
integrable model by using the thermodynamic Bethe ansatz
\cite{YY}. This works by deriving the Bethe equations, which are
coupled polynomial equations (of number of particles in the system).
The energy levels are given by particular sums over the solutions.
The Bethe equations are not solvable in closed form, but if one can
find all the different types of solutions (called ``strings'') in the
continuum limit, the free energy can be computed by calculating the
densities of each of these different types. For an $S$ matrix with
particles in the vector of $O(n)$, the Bethe equations are known
\cite{RW}. The types of solutions are known
for the more general quantum-group algebra $U_q(O(n))$, described
in the language of a lattice ``RSOS''
model \cite{BR,KNS}.  The Boltzmann weights of the RSOS model are
precisely the $S$ matrix of the perturbed $O(n)_k/O(n$--$1)_k$ coset
(up to an overall function which makes $S^\dagger S=1$; this factor
was worked out in \cite{Vays}), and the problem of finding the free
energy is closely related  (an analogous computation and
references are in \cite{FI}). We then can take the $k\to\infty$ limit to
obtain the free energy of the sigma model.

We first discuss $n$ even, where $O(n)$ is simply laced. The function
$\epsilon_0(\theta)$ is defined so that the filling fraction of
particles at rapidity $\theta$ and temperature $T$ is
$1/(1+\exp(\epsilon_0(\theta)))$ (the filling fraction is the density
of particles divided by the density of states). Equivalently,
$T\epsilon_0(\theta)$ is the energy it takes to create a particle of
energy $m\cosh\theta$ over the Fermi sea; the mass $m$ of a particle
is related to the sigma model coupling (the radius of the $n$--$1$
dimensional sphere) in \cite{Hasen}. We also define a set of ``magnon
energies'' $\epsilon_r^{(a)}(\theta)$, where $r=1\dots k-1$ and
$a=1\dots n/2$. The functionals $A^{(a)}_r (\theta)$ are defined as
$$A^{(a)}_r (\theta) \equiv \int_{-\infty}^{\infty} {d\theta'\over 2\pi}
 {n-2\over 2
\cosh[(n-2)(\theta-\theta')/2]} \ln(1+e^{\epsilon_r^{a}(\theta')})$$
while $\widetilde{A}^{(a)}_r(\theta)$ is defined with $\epsilon\to
-\epsilon$.  The matrix $I_{ab}$ is $2-C_{ab}$, where $C_{ab}$ is the
Cartan matrix for $O(n)$, while the matrix $\widetilde I_{rs}=\delta_{r,s-1} +
\delta_{r,s+1}$.
Finally, with ${\cal M}_{ab}(x) = 2\cosh(\pi x/(n-2))\delta_{ab}-I_{ab}$
and ${\cal M}^{-1}$ its matrix inverse,
the function $f_a(\theta)$ is defined as the Fourier transform
$$f_a(\theta) = \int_{-\infty}^{\infty} {dx\over 2\pi}
 e^{ix\theta}({\cal M}^{-1})_{1a}(x)$$
Then an extension of the results of \cite{BR} yields the following
integral equations, valid for even $n> 4$ and $k\ge 2$:
\begin{eqnarray}
\nonumber
\epsilon_0(\theta) &=& {m\over T}\cosh\theta -
 \sum_{a=1}^{n/2} \int_{-\infty}^\infty
d\theta' f_a(\theta-\theta') \ln(1+e^{-\epsilon_1^{(a)}(\theta')})
\\ 
\epsilon_r^{(a)} &=& -\widetilde{A}_0\, \delta_{r,1}\delta_{a,1}
- \sum_{s=1}^{k-1} \widetilde I_{rs} \widetilde A_s^{(a)} 
+ \sum_{b=1}^{n/2}  I_{ab} A_r^{(b)}
\label{ontba}
\end{eqnarray}
The free energy per unit length is then given by
\begin{equation}
F=-{mT\over 2\pi}\int_{-\infty}^{\infty} d\theta \cosh\theta\
\ln(1+e^{-\epsilon_0(\theta)}).
\label{free}
\end{equation}

When $n$ is odd so $O(n)$ is not simply laced, the structure
is more complicated.  Nevertheless, the conjecture still is valid
and the free energy follows from
\cite{KNS}. The function ${\cal M}_{ab}$ is given by their equation (B.10),
while the second equation in (\ref{ontba}) is replaced by their (B.4a)
with its left-hand side replaced by $-A_0\delta_{r,1}\delta_{a,1}$
(note also that the range of $r$ depends on the value of $(a)$).

These equations are straightforward to solve numerically.
The free energy as $m\to 0$
gives the correct value, proportional to the central charge 
$c=k(n-1)(2k+n-4)/(2(k+n-3)(k+n-2))$ of the
$O(n)_k/O(n$--$1)_k$ conformal field theory. The equations remain
well-defined as $k\to\infty$; an infinite number of magnons is
a generic characteristic of models with Lie algebra
symmetries (as opposed to quantum-group structure).

We think the above arguments are convincing for integrable models, but
other cases remain mostly unexplored. The $CP^{n-1}$ sigma
model, which has $G=SU(n)$ and $H=SU(n$--$1)\times U(1)$, is
particularly interesting. This is believed to be not
integrable except for $n$=$2$, where $SU(2)/U(1) \approx O(3)/O(2)$.
The (not conclusive) evidence against integrability is that no local
conserved charges have been found \cite{Gold}, and that anomalies
appear in the non-local conservation laws \cite{Abdalla}.  Our
conjecture may provide a useful way of exploring the model's
properties. The $SU(n)_k/SU(n$--$1)_k\times U(1)$ coset model is dual
to the ``$W^{(k)}$ minimal model'' $SU(k)_{n-1} \times SU(k)_{1}/SU(k)_n$.
In the latter, the perturbing operator ${\cal O}_\sigma$ is denoted
$(k,\bar k;1)+(\bar k,k;1)$.
%, where $k$ is the fundamental representation of $SU(k)$. 
For $k$=$2$, this model is the
$\Phi_{21}$ perturbation of the $n$th minimal model.  Both this
\cite{Zam} and the $k$=$3$ case \cite{Vays} are integrable,
but the counting argument used to prove integrability for $k$=$2,3$
does not yield a conserved current for $k>3$.  However, (at least to first
order in perturbation theory),
all these models have a nonlocal symmetry generated by the chiral
part of the $W^{(k)}_{-1}(1,1;adjoint)$ operator ($\Phi_{15}$ for $k=2$).

There are a number of prospective uses of our conjecture in the
$CP^{n-1}$ model. One could use the truncated conformal scaling
approach \cite{Yurov} to find the low-lying energy levels of the
theory; a signal of integrability is that the levels can cross as the
strength of the perturbation is varied. Also, the conjecture implies
the existence of non-local conserved quantities in the sigma model, by
taking the $k\to\infty$ limit of those in the perturbed coset
model. These do not seem to be anomalous like the ones discussed in
\cite{Abdalla}, so even if $CP^{n-1}$ is not integrable, it still
should have an interesting symmetry structure.

We have found a broadly-applicable and useful feature of $G/H$ sigma
models, a feature which we have conjectured to be completely
general. In fact, we believe it is even true when $G/H$ is not a
symmetric space; the complication is that there are multiple coupling
constants in the sigma model, and multiple perturbations of the coset
model.  Moreover, simple extensions of this conjecture allow for
topological or Wess-Zumino terms in the sigma model action, and also
to supersymmetric sigma models.

\bigskip

I thank P. Arnold, K. Intriligator, Z. Maassarani and
H. Saleur for helpful conversations. I also thank
J.~Ba log and A.~Hegedus for pointing out several typos
in the TBA equations appearing in v1 and in the published version.
This work was supported by a DOE
OJI Award, a Sloan Foundation Fellowship,
and by NSF grant DMR-9802813.

\end{document}